**Translating Electrocardiograms to Cardiac Magnetic Resonance Imaging Useful for Cardiac Assessment and Disease Screening: A Multi-Center Study**

Zhengyao Ding[1*], Ziyu Li[1*], Yujian Hu[2*], Youyao Xu[5], Chengchen Zhao[6], Yiheng Mao[1], Haitao Li[1], Zhikang Li[1], Qian Li,[4] Jing Wang[3], Yue Chen[3], Mengjia Chen[3], Longbo Wang[3], Xuesen Chu[7], Weichao Pan[8], Ziyi Liu[8], Fei Wu[1, #], Hongkun Zhang[2, #], Ting Chen[3, #], Zhengxing Huang[1, #]

[1] College of Computer Science and Technology, Zhejiang University, Hangzhou, China

[2] Department of Vascular Surgery, The First Affiliated Hospital of Zhejiang University School of Medicine, Hangzhou, China

[3] Department of Cardiology, The First Affiliated Hospital, Zhejiang University School of Medicine, Hangzhou, Zhejiang Province, China

[4] Department of Radiology, The First Affiliated Hospital, Zhejiang University School of Medicine, Hangzhou, Zhejiang Province, China

[5] Department of Vascular Surgery, Quzhou People's Hospital, Quzhou, China

[6] Department of Cardiology, The Second Affiliated Hospital of Zhejiang University School of Medicine, Hangzhou, China

[7] China Ship Scientific Research Center, Wuxi, China

[8] Guangdong Transtek Medical Electronics Co., Ltd., Zhongshan, China

# Corresponding authors

* These authors contributed equally to this work.

## Abstract

### Background

Cardiovascular diseases (CVDs), the leading cause of global mortality, require early and accurate diagnosis to improve outcomes. Cardiac magnetic resonance imaging (CMR) provides gold-standard functional and structural insights but remains inaccessible due to cost and complexity. Electrocardiography (ECG) is widely available but lacks CMR's granularity. We propose **CardioNets**, a deep learning framework that translates 12-lead ECG data into functional parameters and synthetic images closely aligned with CMR imaging, enabling scalable, low-cost cardiac assessment.

### Methods

CardiacNets is built upon cross-modal contrastive learning to align ECG signals with CMR-derived cardiac phenotypes (e.g., ejection fraction, myocardial mass). The framework operates in two phases: (1) Cross-modal alignment, where ECG is aligned with CMR functional data, and (2) CMR image synthesis, where a masked autoregressive model generates high-fidelity CMR images from ECG input. The model was trained and validated on 159,819 samples from four datasets, including the UK Biobank (n=42,483) and MIMIC-IV-ECG (n=164,550), with external testing on two independent clinical datasets (n=3,767). Tasks included multi-disease screening (e.g., cardiomyopathy, coronary artery disease) and cardiac function assessment.

### Findings

In the UK Biobank, CardioNets improved cardiac phenotypes regression $R^2$ by 28.1% over model trained from scratch and 15.2% over self-supervised pretraining, and boosted cardiomyopathy screening AUC by 38.2% and 5.5%, respectively. In MIMIC, it raised AUC for pulmonary hypertension screening by 3.0% over model trained from SSL. Synthesized CMR images by CardioNets showed 65.0% higher SSIM and 15.5% higher PSNR than prior method. In the reader study, cardiomyopathy screening using CardioNets (ECG only) achieved a 15.2% higher accuracy compared to the average performance of human physicians using both ECG and real CMR. In most disease-screening tasks, CardioNets performance closely matched CMR-based models, demonstrating ECG's potential as a low-cost surrogate for CMR in cardiovascular screening.

### Interpretation

CardioNets bridges the accessibility gap between ECG and CMR, offering a cost-effective solution for population-scale screening. By translating ECG into CMR-level insights, CardioNets bridges gaps in accessibility and interpretability, enabling population-scale screening without compromising diagnostic accuracy. The model's ability to generate CMR images from ECG enhances clinical decision-making, particularly in resource-limited settings. Future work should validate real-world implementation and regulatory pathways for AI-generated imaging.

## Research in Context

### Evidence before this study

Cardiovascular diseases (CVDs) are the leading cause of global mortality, yet low-cost, large-scale screening remains a challenge due to the limitations of current diagnostic tools. Electrocardiography (ECG) is cost-effective, widely accessible, and commonly used in clinical settings. However, it primarily detects electrical abnormalities and is unable to evaluate structural or functional heart issues. In contrast, cardiac magnetic resonance imaging (CMR) offers a comprehensive assessment of heart structure and function, but its high cost, complexity, and limited availability make it inaccessible, particularly in resource-limited settings. Although few prior studies have attempted to integrate ECG and CMR, they struggled to effectively identify structural heart diseases and lacked the capacity to generate high-fidelity CMR images. The potential for integrating ECG with CMR through cross-modal artificial intelligence (AI) remains an underexplored area of research.

### Added value of this study

This study introduces CardioNets, a new cross-modal AI framework that translates standard 12-lead ECG data into diagnostic CMR-derived functional parameters and synthetic CMR images. By leveraging contrastive learning, CardioNets aligns ECG waveforms with CMR biomarkers, such as ejection fraction and global longitudinal strain. Additionally, it employs a masked autoregressive model to generate high-quality CMR images from ECG data. Trained on extensive datasets like the UK Biobank and MIMIC-IV, CardioNets significantly enhances ECG's ability to detect CVDs, with performance comparable to CMR-based diagnostic model. External reader study showed a 15.2% accuracy gain with CardioNets over physicians using ECG and real CMR.

### Implications of all available evidence

CardioNets bridges the accessibility gap between ECG and CMR, offering a cost-effective solution for early CVD detection in resource-limited regions. The synthetic CMR images generated by CardioNets enhance interpretability, facilitating informed decision-making without the need for specialized imaging. From a policy perspective, this AI-driven model offers a cost-effective and scalable solution for population-level screening of non-communicable diseases, aligning with global health objectives. In the long term, CardiacNets has the potential to complement existing diagnostic pathways, assist in optimizing CMR referral strategies, and provide a framework for the application of AI in other resource-constrained diagnostic domains.

## Introduction

Cardiovascular diseases (CVDs) remain the leading cause of mortality worldwide, with their share of global deaths increasing significantly in recent years[1-3]. Among these, the mortality rate associated with cardiomyopathy has risen by 7.6%, significantly contributing to the growing burden of cardiovascular disease[4]. The clinical presentation of CVDs is often complex and multifactorial, leading to missed or misdiagnosed cases, which can delay treatment and compromise patient care[5,6]. To effectively assess cardiovascular health, clinicians utilize a variety of diagnostic modalities, including electrocardiography (ECG)[7] and cardiovascular magnetic resonance imaging (CMR)[8]. ECG measures the electrical activity of the heart, helping identify conditions like myocardial infarction and arrhythmias. It is widely adopted due to its non-invasive nature, low cost, and accessibility across healthcare settings. However, while ECG offers valuable insights into electrical abnormalities, it has limitations in assessing structural and functional cardiac conditions[9]. On the other hand, CMR provides a more comprehensive evaluation, offering detailed imaging of cardiac morphology, function, myocardial perfusion, and tissue characterization[10-13], making it especially valuable in assessing structural heart diseases like cardiomyopathy[14-16]. Despite its strengths, CMR's high costs and operational complexity limit its accessibility, particularly in resource-constrained settings[17]. This discrepancy highlights the urgent need for methods that can enhance ECG's diagnostic power, particularly as a preliminary tool for CVDs opportunistic screening.

Recent advancements in large-scale multimodal datasets, such as the UK Biobank (UKB) and MIMIC[18,19], coupled with breakthroughs in deep learning, offer a transformative opportunity to address this gap. Cross-modal contrastive learning approaches, exemplified by CLIP (Contrastive Language-Image Pre-training)[20], have demonstrated the ability to align diverse data modalities and enable high-precision tasks like zero-shot classification, where models can make predictions without task-specific training. Additionally, generative models, such as autoregressive model[21,22] and stable diffusion model[23], have harnessed CLIP's text encoder to guide the generation of high-fidelity images from textual descriptions, further expanding the potential of cross-modal models in healthcare. These advances pave the way for integrating ECG and CMR in innovative ways that could improve disease detection and treatment planning[24-26].

In response to these challenges and opportunities, we introduce **CardioNets**, a novel approach that utilizes cross-modal alignment and generative model to enhance cardiac assessment (Fig. 1). Unlike traditional methods, which align text-image pairs of similar content[20], CardioNets focuses on strengthening the relationship between a more robust modality (CMR) and a weaker modality (ECG) to improve the latter's performance. While previous studies have explored the joint analysis of ECG and CMR data[24], focusing primarily on broad cardiovascular state representations, our work centers specifically on CVDs and is explicitly designed for large-scale, opportunistic screening—a clinically impactful application highlighted in recent population-level AI screening studies[27]. This approach enhances the ECG data, leading to more precise cardiac assessments. Furthermore, CardioNets integrates a generative model[28] to produce dynamic CMR sequences from ECG data, thereby improving interpretability and supporting clinical decision-making.

We evaluated CardioNets in two phases: (1) cross-modal learning and ECG-based CMR image generation, and (2) downstream cardiac assessments, including the prediction of cardiac phenotypes and screening for coronary artery disease, heart failure, cardiomyopathy, and pulmonary arterial hypertension. In the UK Biobank, CardioNets improved the regression $R^2$ for cardiac phenotypes by 28.1% compared to models trained from scratch and by 15.2% over models trained with self-supervised pretraining. It also enhanced cardiomyopathy screening AUC by 38.2% and 5.5%, respectively. In the MIMIC dataset, CardioNets boosted AUC for pulmonary hypertension screening by 3.0% over models trained with self-supervised learning (SSL). The synthesized CMR images generated by CardioNets showed a 65.0% increase in SSIM and a 15.5% improvement in PSNR compared to previous method[24]. A reader study further demonstrated the practical utility of CardioNets, particularly in assisting both resident and attending physicians in cardiomyopathy screening. In conclusion, CardioNets offers a scalable, cost-effective solution for early CVD detection, with the potential to transform clinical practice, particularly in resource-limited settings with restricted access to advanced imaging technologies.

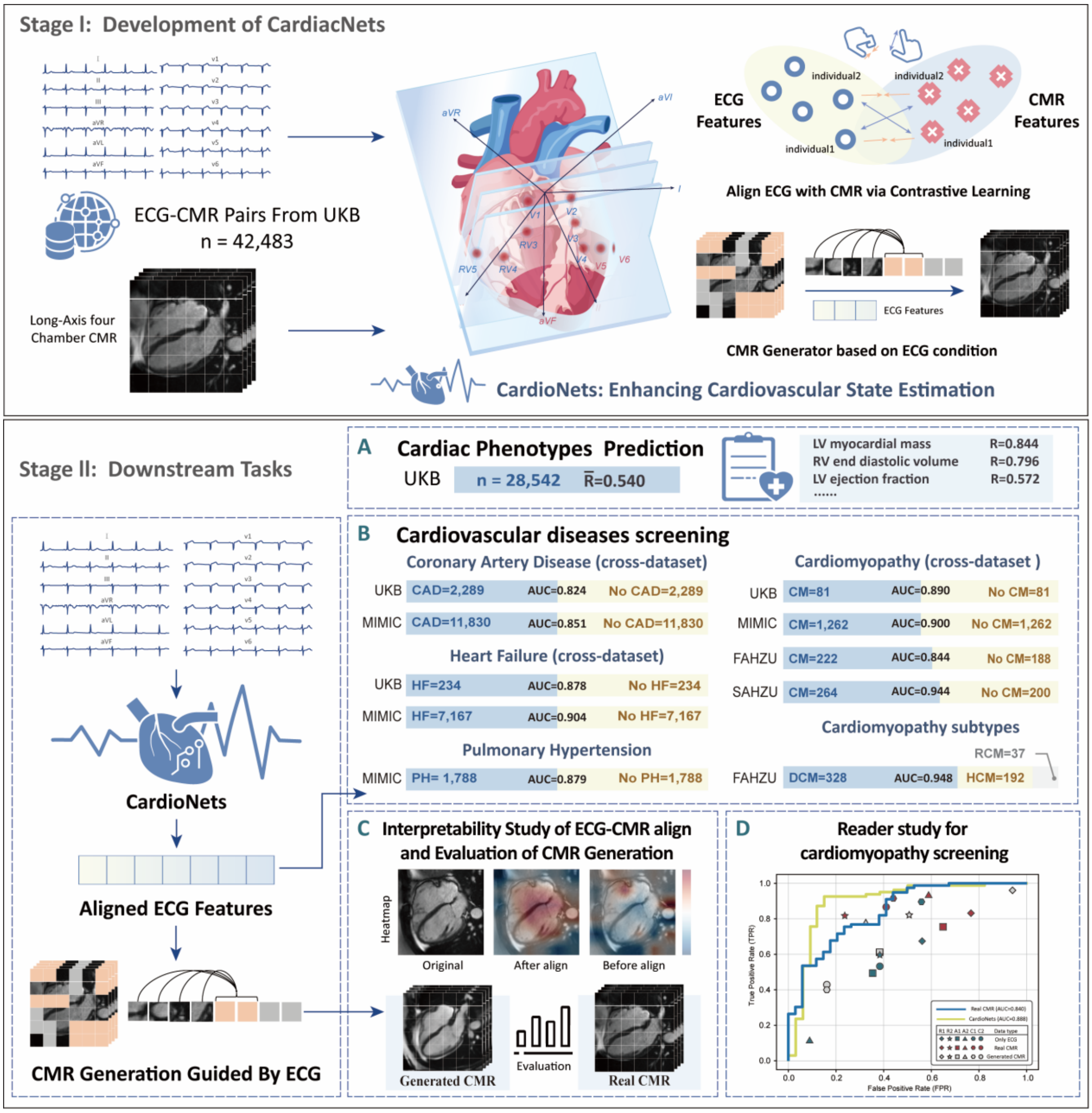

**Fig. 1: Overview of the study design. Stage I:** Development of CardioNets. Using the UKB dataset, we trained two core components of CardioNets: (1) an ECG model trained with ECG-CMR contrastive learning

to capture relevant information from CMR, and (2) a CMR generation model trained with a masked autoregressive model conditioned on ECG. **Stage II:** Downstream tasks. The datasets used are from UKB, MIMIC-IV-ECG, The First Affiliated Hospital of Zhejiang University School of Medicine (FAHZU), The Second Affiliated Hospital of Zhejiang University School of Medicine (SAHZU). All numbers annotated in the figure are represent the total sample sizes of the datasets. All tasks used only ECG data to validate the model's performance, including: (A) Prediction of cardiac phenotypes, (B) Screening for cardiovascular diseases, particularly cardiomyopathy, (C) Interpretability study of ECG-CMR align and evaluation of CMR generation, and (D) Reader study for cardiomyopathy screening.

## Methods

### Dataset for Model Development and Evaluation

An overview of the datasets used for model development, validation, and external evaluation is shown in Extended Fig. 1. Below, we provide a detailed description of the datasets used for model development and validation.

For model pretraining, 42,823 ECG–CMR pairs were sourced from the UK Biobank (UKB)[18], a population-based cohort of approximately 500,000 participants aged 40–69, recruited across the UK since 2006. Imaging data were collected during the initial imaging visit (instance 2), which ensured the pairing between ECG and CMR[29]. These pairs were used to train a contrastive learning model with the CMR encoder frozen. The CMR encoder had previously undergone self-supervised pretrained on 194,664 CMR images, which included a mixture of phenotype-guided synthetic CMRs and real CMRs, as detailed in our previous work[30]. This pretraining enabled the extraction of rich semantic representations. The resulting ECG encoder is referred to as the "aligned ECG encoder". In the development of the ECG-conditioned CMR generative model, the same 42,823 UKB pairs were used for both training and validation, with features from the frozen aligned ECG encoder serving as conditioning input. Unless otherwise specified, downstream results reported in this study were obtained by fine-tuning the aligned ECG encoder.

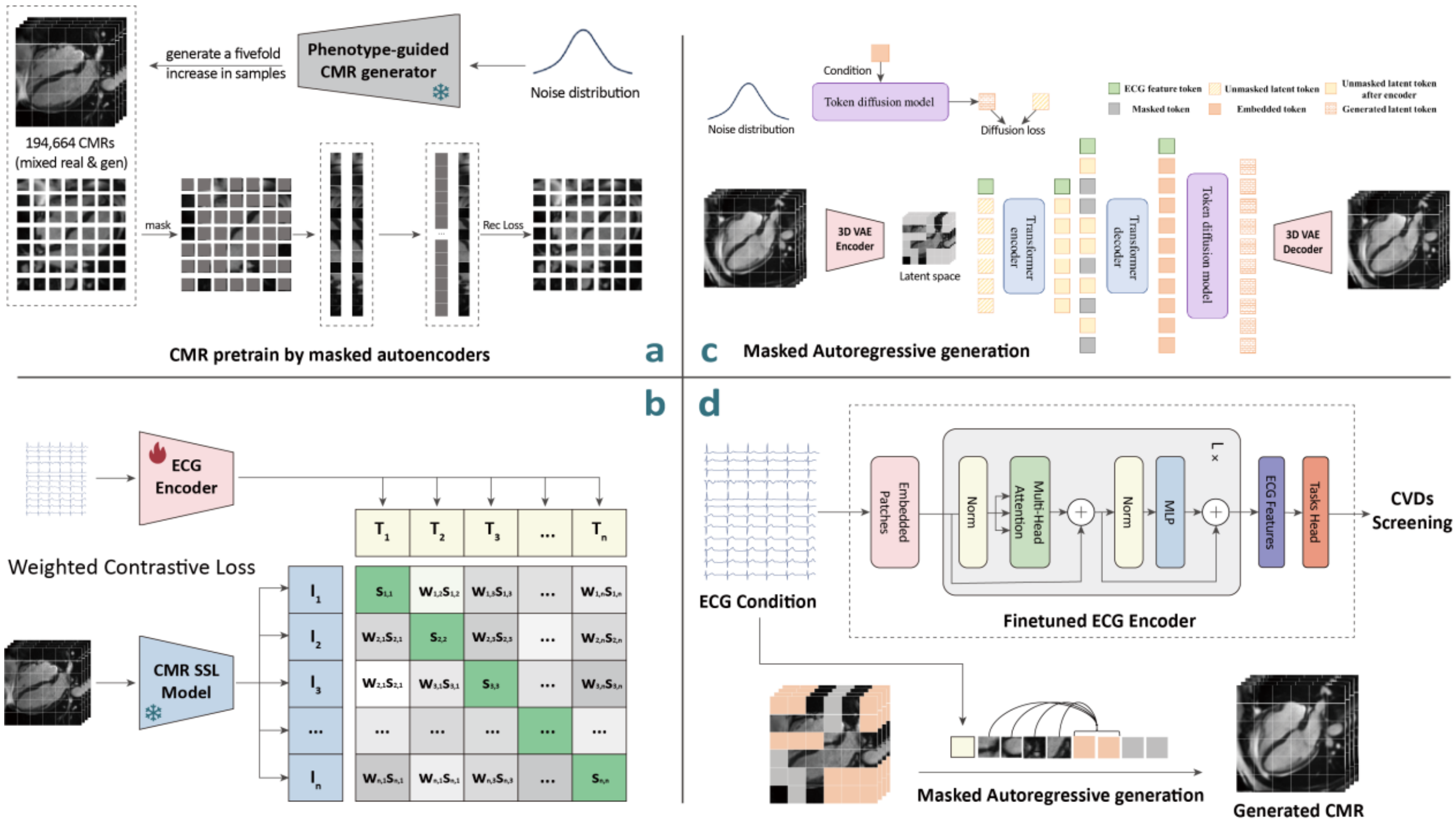

**Fig. 2: Model diagram. a.** CMR self-supervised model, leveraging masked self-supervised learning and optimized to minimize reconstruction loss. To enrich pretraining, we incorporated a phenotype-guided CMR generation approach (from our previous work[30]), which expanded the dataset with mixed real and generated CMRs, thereby enabling the model to learn more robust CMR representations. **b.** ECG-CMR contrastive learning, with the parameters of the CMR self-supervised model in subgraph a frozen, optimized to minimize the weighted contrastive loss, where $w_{i,j}$ denotes the phenotype-based weight and $s_{i,j}$ denotes the cosine similarity between embeddings. **c.** ECG-guided CMR image generation, using the pre-trained ECG encoder in subgraph b as the feature extractor. **d.** Fine-tuning the aligned ECG encoder to adapt to various downstream tasks and use pretrained generative model in subgraph c generate the corresponding CMR images.

For our downstream analyses, we examined multiple tasks across several datasets, including the UKB, MIMIC-IV-ECG[31], and private datasets from two hospitals: the First and Second Affiliated Hospitals of Zhejiang University School of Medicine (FAHZU, SAHZU):

- **UKB analyses**: (i) Phenotype regression was conducted on a subset of 28,542 participants from the original 42,823 ECG–CMR pairs, all of whom had complete data on ECG, CMR, and 82 cardiac indices (Supplementary Table 1). These indices were derived from CMR according to UKB Category 157[32]. (ii) CVD screening encompassed CAD (n=2,289), HF (n=234), and CM (n=81), with negatives sampled drawn at varying ratios of 1:1, 1:2, 1:5, and 1:10. Diagnostic labels were assigned based on ICD codes (field ID 41270), with the corresponding first in-patient diagnosis dates retrieved from field ID 41280. The dates of ECG and CMR acquisition were determined from the "Date of attending assessment centre" (field ID 53). Only diagnoses with dates earlier than or equal to the imaging date were retained, ensuring temporal alignment between labels and inputs, as chronic cardiovascular diseases progress slowly with relatively stable phenotypes[33-35]. We employed five-fold cross-validation and reported mean ± standard deviation performance across folds.

- **MIMIC-IV-ECG analyses**: From approximately 160,000 participant records, we constructed cohorts for CAD (n=11,830), HF (n=7,167), CM (n=1,262), and PH (n=1,788), with negatives samples drawn using the same ratios as in the UKB. Disease labels were obtained from the curated MIMIC-IV-ECG-Ext-ICD extension[36], which links ECG recordings to hospital or emergency admissions and derives diagnostic codes from discharge records (standardized to ICD-10). This ensured that the disease labels corresponded accurately to the ECG inputs. We applied a 60:20:20 train/validation/test split and reported results on the held-out test set. Since MIMIC does not include CMR imaging, cardiac structural phenotype regression was not performed.

- **FAHZU analyses**: Dataset 1 (ECG-only) comprised 1,705 cases of DCM, 925 cases of HCM, and 162 cases of RCM, used for three-class cardiomyopathy classification and evaluated through five-fold cross-validation. Dataset 2 (ECG-only) included 222 cardiomyopathy patients and 188 controls, and was employed for external validation. Dataset 3 (paired ECG–CMR) consisted of 77 cardiomyopathy patients and 34 controls, used in a reader study involving two attending physicians, two residents, and two associate chief physicians. Disease labels were derived from electronic health record discharge diagnoses.

- **SAHZU analyses**: Contained 264 cardiomyopathy patients and 200 controls, used for external validation.

Across all datasets, ECGs were standardized to the same format as the UKB (12-lead, 10 seconds, 500 Hz). Pulmonary hypertension cases were excluded from the UKB due to their low prevalence. For both the UKB and MIMIC datasets, disease labels were assigned based on ICD-10 codes: Coronary Artery Disease (CAD: I20–I25), Cardiomyopathy (CM: I42–I43), Heart Failure (HF: I50), and Pulmonary Hypertension (PH: I27). For the private hospital datasets (FAHZU and SAHZU), cardiomyopathy labels were obtained from discharge diagnoses recorded in electronic health records. Data distributions are summarized in Supplementary Tables 2–4. Ethical approval was granted by the institutional review board of FAHZU (Approval No. IIT20240507B).

**Data processing and augmentation**

For ECG data, we followed the pre-processing and augmentation strategy outlined by Na et al[32]. Specifically, all signals were resampled to 250 Hz, randomly cropped during training, and passed through a band-pass filter (0.67–40 Hz) to suppress baseline wander and high-frequency noise. Standardization was then applied before converting the signals to tensors. During validation and testing, random cropping was replaced with fixed multiple crops to ensure stable evaluation. For CMR data, no geometric or spatial augmentations were applied. All images were simply normalized to the range of $(-1, 1)$ for both training and validation.

**Self-supervised method for CMR**

We employed SSL on CMR images to derive robust representations (Fig. 2a), which then facilitated contrastive ECG–CMR pretraining. To increase the diversity and scale of the training data, we integrated phenotype-guided generative augmentation using the CPGG framework[30] developed in our prior work. Specifically, CPGG was trained on 32,444 four-chamber cine CMR sequences from the UK Biobank, where 82 cardiac phenotypes were fully available for generation. The dataset was split into 25,955 samples for training, 3,244 for validation, and 3,245 for testing. Using this cohort, CPGG generated a five-fold synthetic dataset, which was merged with the real CMR data, creating an expanded corpus of 194,664 samples for self-supervised pretraining. In our prior work, we have demonstrated that this strategy of combining large-scale phenotype-conditioned generation with SSL yields robust and semantically meaningful CMR representations.

Masked self-supervised learning[37] was applied to this mixed dataset (real + synthetic, 500% mix). A Vision Transformer (ViT) with a patch size of $16\times16$ served as the backbone. The temporal dimension of the long-axis cine CMR (50 frames) was treated as the channel dimension, and random masking was applied across channels at a ratio of 75%. The model was trained to reconstruct the masked inputs using a reconstruction loss. Training was conducted on an 80 GB A800 GPU with a batch size of 32 for 400 epochs, including 40 warm-up epochs where the learning rate was linearly increased from 0 to $1\times10^{-4}$.

Importantly, during the subsequent contrastive ECG–CMR training, the CMR encoder was kept frozen. This pretrained encoder, enhanced with phenotype-conditioned generative augmentation via CPGG, embedded high-level cardiac semantic information into the learned representation. This approach fundamentally differs from previous methods that jointly trained ECG and CMR encoders.

**Contrastive ECG-CMR Pre-training**

Building on the pretrained CMR encoder (with frozen parameters), we initialized the ECG encoder using weights from the ST-Mem self-supervised model[38].The objective was to align ECG and CMR features within a shared latent space through contrastive learning (Fig. 2b).

A key challenge in contrastive pre-training is that the standard InfoNCE loss treats all non-paired samples as negatives, even when they share highly similar clinical characteristics (e.g., two normal ECG–CMR pairs). To address this, we implemented a weighted InfoNCE objective that down-weights negatives corresponding to clinically similar samples in the CMR semantic space.

Specifically, the frozen CMR encoder (pretrained with phenotype-guided augmentation as described above) was used to extract per-sample CLS embeddings $h \in R^{768}$. These embeddings were then reduced using PCA (fitted only on the training set) to $\phi \in R^{128}$, which served as continuous pseudo-phenotypes encoding CMR semantics.

For ECG embeddings $z_e$ and CMR embeddings $z_c$, cosine similarity was computed as $s_{ik} = cos(z_{e,i}, z_{c,k})$. A weight matrix $W \in R^{\boldsymbol{B}\times\boldsymbol{B}}$ was constructed from pairwise distances in the CMR-PCA space:

$$W_{ik} = 1 - exp\left(-\frac{\| \phi_i - \phi_k \|_2^2}{\sigma^2}\right), W_{ii} = 0. \tag{1}$$

Thus, negatives that are close in CMR semantic space (clinically similar) receive smaller weights, contributing less to the denominator in the InfoNCE loss. The per-direction weighted InfoNCE loss is:

$$\mathcal{L}_{e\to c} = -\frac{1}{B}\sum_{i=1}^{B} log\frac{exp(s_{ii}/\tau)}{exp(s_{ii}/\tau) + \sum_{k\neq i} W_{ik} exp(s_{ik}/\tau)}, \tag{2}$$

with a symmetric objective:

$$\mathcal{L} = \frac{1}{2}(\mathcal{L}_{e\to c} + \mathcal{L}_{c\to e}). \tag{3}$$

The model was trained on an 80GB A800 GPU with a batch size of 256 for up to 400 epochs. The first 40 epochs were dedicated to learning rate warm-up, gradually increasing the learning rate from 0 to $1\times10^{-3}$. The model checkpoint with the lowest validation loss was saved for use in downstream tasks.

This weighted contrastive strategy preserved the benefits of instance-level alignment while mitigating the risk of overfitting to individual samples, thereby encouraging the extraction of shared disease-relevant features across clinically similar ECG–CMR pairs.

**ECG2CMR generative model**

Similar to Stable Diffusion[23], our model performs generation in the latent space. We employ a 3D Variational Autoencoder (3D-VAE) as the compressor for CMR data, with spatial and temporal down-sampling rates denoted as $f_s$ and $f_t$, respectively. For a long-axis cine CMR input of shape $(1, T, H, W)$ ,

the resulting latent representation $x$ has dimensions of $|x| \times \frac{T}{f_t} \times \frac{H}{f_s} \times \frac{W}{f_s}$, where $|x|$ denotes the latent dimensionality.

For ECG-to-CMR generation, we built upon the Masked Autoregressive Model (MAR)[22]. The MAR is a variant of the standard autoregressive model that enables simultaneous prediction of multiple tokens conditioned on the observed ones, offering improved efficiency and flexibility over conventional step-by-step autoregressive approaches. We used a Transformer network with bidirectional attention as the backbone of the autoregressive model. Specifically, the compressed CMR representations generated by the 3D-VAE encoder are divided into non-overlapping tokens of shape $|x| \times p_t \times p_s \times p_s$, where $p_t \times p_s$ are temporal and spatial strides, respectively. A dynamic masking ratio was applied during training, with masked tokens reconstructed using a diffusion-based denoising loss. The complete mathematical formulation of the autoregressive and diffusion objectives is provided in Supplementary Methods.

Next, we encoded the input ECG using the aligned ECG encoder to obtain ECG feature representations. These features were then prepended to the CMR token sequence as a special conditional token (Fig. 2c). A bidirectional attention mechanism was applied to this combined sequence, enabling each CMR token to integrate conditional information from the ECG, thus facilitating ECG-guided CMR generation. For generation, we adopted an iterative decoding strategy[39] to synthesize the cine CMR sequence. The process began with an empty latent representation of the CMR, where all tokens were initially masked. Iterative decoding was performed over $K$ steps, where at each step, the model predicted a subset of the remaining masked tokens. A portion of the predicted tokens was randomly retained, while the masking ratio followed a cosine schedule across iterations. This progressive refinement ensures that the model gradually improves the quality of the generated CMR representations.

In our implementation, the 3D-VAE is configured with a latent dimensionality of 16, $f_t$ of 2, and $f_s$ of 8. For the masked autoregressive model, we employed a 12-layer encoder–decoder architecture with a latent embedding size of 768 and a patch size of 5×2×2. This patch size was chosen to reduce computational load, by leveraging the inherent temporal redundancy of cine CMR data. During training, the masking ratio was randomly sampled between 0.7 and 1.0. For the diffusion component, we adopted a cosine-based noise schedule, using 1,000 timesteps during training and 100 timesteps during inference. The noise prediction network consisted of a 3-block multilayer perceptron (MLP) with a channel width of 1,024. The iterative decoding process was performed over 16 steps to progressively reconstruct the CMR sequence. Experiments were conducted on an NVIDIA A800 GPU (80 GB). The models were trained using the AdamW optimizer with a learning rate of $8\times10^{-4}$ for 400 epochs.

**Downstream tasks**

For downstream tasks, we used only the ECG encoder from the Contrastive ECG-CMR Pre-training model, excluding the CMR encoder (Fig. 2d). The ECG encoder extracts high-level features from ECG signals and embeds CMR information. The pre-trained autoregressive generative model was then used to generate high-fidelity CMR images with authentic cardiac details, offering valuable insights for clinicians. The objective was to generate classification outputs aligned with CVD labels or regression outputs that closely match the ground truth for 82 cardiac phenotypes. Training was conducted with a batch size of 16 for up to 100 epochs, using a learning rate of $5\times10^{-5}$ with cosine annealing scheduling, a 10-epoch warm-up, and early stopping.

To comprehensively evaluate CardioNets, we compared it with several baseline methods using the same ECG model architecture. These methods included SSL ECG (initialized with pretrained weights from Na et al.[38] and fine-tuned on each downstream task), and Phen_sup ECG (first pretrained on the UK Biobank dataset for regression of 82 cardiac phenotypes, then fine-tuned on downstream tasks, except phenotype regression), Scratch ECG (randomly initialized and trained in a fully supervised manner), and CMR Ref (when CMR data were available, initialized with weights from our self-supervised pretrained CMR encoder). CardioNets was initialized with the ECG encoder from the Contrastive ECG–CMR Pre-training model and fine-tuned under the same training setup. For datasets used solely for direct inference, we adopted models trained on the MIMIC dataset, which provides a larger cohort than UKB, and directly applied them for evaluation across baselines (CardioNets, SSL ECG, Phen_sup ECG, and Scratch ECG).

### Quantitative assessment and statistical analysis

For classification tasks, we employed several evaluation metrics: area under the curve (AUC), accuracy, sensitivity, specificity, positive predictive value (PPV), and negative predictive value (NPV). In our three-class classification, we binarized the labels using a one-vs-rest approach prior to evaluation. For regression tasks related to cardiac phenotypes, we used the coefficient of determination ($R^2$), Pearson correlation coefficient, mean absolute error (MAE), and root mean squared error (RMSE) as the primary evaluation metrics. For the MIMIC dataset and the FAHZU and SAHZU external inference datasets, 95% confidence intervals (CIs) were computed using the Wilson Score Interval[40] for all metrics except AUC, for which CIs were estimated via bootstrap resampling (n = 1000). For the UKB and FAHZU multiclass classification datasets, we reported the mean and std based on five-fold cross-validation.

To compare CardioNets with baseline methods (model-based CMR, SSL ECG model, phen_sup ECG model, and Scratch ECG model), we evaluated AUC, balanced accuracy, and mean average precision (mAP). Statistical significance was assessed using the DeLong test for AUC, and permutation tests (N = 10,000) for balanced accuracy and mAP. In addition, 95% confidence intervals were estimated using bootstrap resampling with 10,000 iterations. The statistical tests report *p*-values, confidence intervals, and effect sizes, with effect sizes quantified using Cohen’s *d*. For the five-fold cross-validation experiments, all validation samples were aggregated into a unified evaluation set. This pooled approach ensures consistent model comparison, although the AUC may differ slightly from the cross-validation mean due to sample reweighting.

### Software for data process and model development

We utilized several libraries for data processing and model development, including numpy (version 1.25.2)[41], sklearn (version 1.1.1)[42], scipy (version 1.11.2)[43], simpleITK (version 2.3.1)[44] and pandas (version 2.2.1). For model development, we employed pytorch (version 1.11.0).

## Results

### Cardiac assessments

We evaluated the performance of the fine-tuned CardioNets model using two public datasets (UKB, MIMIC) and two private datasets (FAHZU, SAHZU) for predicting cardiac structural phenotypes and screening for CVDs (Fig. 3a-b). In the UKB dataset, which provides paired ECG–CMR data, CardioNets was benchmarked against both standalone ECG-based and CMR-based models. In contrast, for MIMIC and the majority of private hospital datasets (FAHZU Datasets 1–2 and SAHZU), no CMR was available such that comparisons were limited to ECG-based models. Of note, FAHZU Dataset 3 contained paired ECG–CMR data and was specifically used for the reader study; results from this experiment are presented separately. Baseline models were defined as described in the Methods section, including ECG-based variants (trained from scratch, initialized with self-supervised learning, or optimized for phenotype regression), as well as a CMR-based model initialized with a self-supervised pretrained encoder.

For cardiac phenotype prediction across 82 indicators (e.g., atrial ejection fractions and chamber volumes), CardioNets achieved an average coefficient of determination ($R^2$) of 0.310, outperforming both ECG-only baselines (0.269 and 0.242), though still trailing behind the CMR-based model (0.511). We also compared CardioNets with two recent state-of-the-art methods, Cross-AE[24] and MMCL[45], and observed that CardioNets significantly outperformed baselines in predicting cardiac phenotypes (Supplementary Fig. 1). More detailed results are provided in Supplementary Fig. 2 and the Supplementary Files (see Supplementary Table 5 for a detailed description of its contents).

For CVD screening, CardioNets demonstrated substantial improvements over ECG-based baselines across three conditions—coronary artery disease (CAD), heart failure, and cardiomyopathy—in the UKB dataset. Notably, for cardiomyopathy detection, CardioNets outperformed the scratch-trained and SSL ECG models by 38.2% and 5.5%, respectively. In CAD screening, CardioNets even surpassed the CMR-based model. In the MIMIC dataset, which included four conditions (CAD, heart failure, cardiomyopathy, and pulmonary arterial hypertension), CardioNets consistently exhibited robust performance. It outperformed the baseline ECG models across all conditions, achieving a 3.4% improvement over the Phen_sup ECG model and a 3.0% improvement over the SSL ECG model in pulmonary arterial hypertension screening.

To further assess model's generalizability, we applied the cardiomyopathy classification model, trained on the MIMIC dataset, to the FAHZU and SAHZU datasets. CardioNets continued to outperform the baseline ECG models in both external datasets, suggesting that the alignment between ECG and CMR enables the ECG encoder to effectively capture the structural information inherent in CMR. This alignment enhances the ECG model's performance across diverse datasets. Comprehensive results, including sensitivity, specificity, accuracy, positive predictive value (PPV), negative predictive value (NPV), and F1-score, are provided in Supplementary Tables 6–11.

Label efficiency refers to the amount of training data and labels required to achieve a target performance level for a specific downstream task, emphasizing the annotation workload for medical experts. CardioNets demonstrated exceptional label efficiency across various disease screening tasks in the MIMIC dataset (Fig. 3c). Remarkably, CardioNets achieved performance comparable to standalone scratch ECG models while utilizing only 10% of the training data, highlighting its potential to mitigate data scarcity. This capability is crucial for real-world applications, as it alleviates the annotation burden on clinical experts, making the model more applicable across diverse clinical settings[46].

The CVDs screening results were obtained by fine-tuning and evaluating models using all available positive samples and a randomly selected, approximately equal number of negative samples. While this setup allows for controlled comparisons, it may not fully reflect real-world clinical conditions. To better assess the performance of CardioNets under more realistic class imbalance conditions, we evaluated the models using varying negative-to-positive sample ratios of 1:1, 2:1, 5:1, and 10:1 (Supplementary Fig. 3). Except for the scratch-trained ECG model, the AUC performance of CardioNets, the SSL ECG model, Phen_sup ECG model, and the CMR reference remained relatively stable across different imbalance ratios. Notably, CardioNets demonstrated performance comparable to the CMR reference and consistently outperformed the baseline ECG models, highlighting its robustness and superior performance in imbalanced data scenarios.

In addition, we present the statistical comparisons of CardioNets with the CMR reference and baseline ECG models in terms of balanced accuracy (BA), mean average precision (mAP), and AUC (Supplementary Figs. 4–6). *P*-values between models were calculated using permutation tests, with N = 10,000 permutations conducted for each pairwise comparison. We also highlight the performance gains of CardioNets over each baseline method, along with the corresponding sample size (n). All statistical significance analyses, including 95% confidence intervals and effect size metrics, are detailed in the Supplementary Files.

**Screening for subtypes of cardiomyopathy**

We evaluated our model's performance on FAHZU Dataset 1 to assess its effectiveness in three-class cardiomyopathy classification (DCM, HCM, and RCM), comprising 2,782 samples as described in Methods. Fig. 4 presents the results of five-fold cross-validation using CardioNets, baseline ECG models, with classification performance shown using a one-vs-rest (OvR) approach. Compared to DCM and HCM, classification performance for RCM was notably lower when using the baseline ECG models, likely due to the limited number of cases. However, CardioNets achieved the greatest improvement in RCM classification, reaching an AUC of 93.3%. Confusion matrix analysis further confirms that CardioNets offers superior discriminative ability for rare cardiomyopathy subtypes. These results underscore the strong generalization ability of CardioNets, particularly in accurately identifying rare cardiomyopathy subtypes with limited training samples.

Comprehensive metrics—including sensitivity, specificity, accuracy, PPV, NPV, and F1-score—are available in Supplementary Tables 12, with all significance analyses detailed in the Supplementary Files. In addition, statistical comparisons between CardioNets and baseline models in terms of balanced accuracy, mAP, and AUC are reported in Supplementary Fig. 7.

**Interpretation study and evaluation of CMR generation**

To explore the interpretability gained through ECG–CMR alignment, we employed the similarity matrix heatmap method from MMCL[45] to compare CardioNets with the baseline ECG models (Fig. 5a). The results demonstrate that, following alignment with CMR, CardioNets showed higher cosine similarity between ECG signals and cardiac regions in the CMR images, while exhibiting lower similarity with peripheral, less relevant areas. In contrast, the similarity matrices for both the pretrained and scratch ECG models displayed more scattered attention patterns, lacking a clear focus. This suggests that CardioNets effectively learned to prioritize cardiac-relevant regions and ignore less meaningful areas for downstream tasks. These findings indicate that CardioNets successfully captures CMR-derived structural information, which guides its downstream predictions. Further interpretability comparisons are provided in the Supplementary Files.

To evaluate the quality of these synthesized CMR images, we report both qualitative and quantitative results (Fig. 5b-c). We observe that CardioNets effectively preserve key physiological structures, such as the myocardium and valves, along with phenotype-related indicators commonly derived from CMR, including left ventricular myocardial mass. To further assess this, a pretrained CMR encoder was used to extract features from both generated and real CMR images in the test set. These features were then visualized and compared using *t*-distributed stochastic neighbour embedding (*t*-SNE), a dimensionality reduction technique, to assess the similarity of their distributions in a lower-dimensional space. As shown, the *t*-SNE plots, histograms, and cumulative distribution function (CDF) plots reveal no substantial differences in feature distributions between generated and real CMR images. Notably, the distributions of generated CMRs display two relatively distinct peaks or clusters compared with the smoother patterns of real CMRs. This clustering is attributed to the generative model's tendency to regress toward the mean during training. While real CMR feature distributions also contain subtle multimodal structures, the generative process amplifies these peaks, making them more pronounced. This phenomenon likely reflects both the intrinsic heterogeneity of cardiac phenotypes and current limitations of generative modelling, where transitional or boundary regions in the phenotype space are less faithfully represented.

To further assess generation quality, we calculated structural similarity index (SSIM), peak signal-to-noise ratio (PSNR), mean absolute error (MAE), and root mean squared error (RMSE) between each synthesized CMR and its paired ground truth. We compared the performance of CardioNets against models based on cross-modal autoencoders[24] and diffusion-based[28] methods. CardioNets consistently outperformed these baselines across all evaluation metrics. For instance, compared with the cross-modal autoencoder, CardioNets achieved a 65.0% higher SSIM and a 15.5% higher PSNR.

In addition to image-level quality metrics, we evaluated the utility of generated CMRs in a downstream phenotype regression task. Specifically, when fed into a pretrained CMR-based model, generated CMRs achieved an $R^2$ of $0.2045 \pm 0.0051$, compared with $0.4417 \pm 0.0035$ for real CMRs and $0.2562 \pm 0.0077$ for direct ECG-to-phenotype prediction (as shown in Supplementary Fig. 8). These results indicate that generated CMRs retain non-trivial clinical information, although some performance is lost relative to direct ECG predictors due to compounded errors in the two-step pipeline.

Beyond these quantitative comparisons, we also assessed the perceptual realism of generated CMRs through a reader study involving six physicians of varying seniority, who were asked to distinguish between real and generated images. The detailed design and results of this human evaluation are presented in the Reader Study section.

The structural and distributional consistency of the generated images, despite some amplification of multimodal clustering, demonstrates that CardioNets captures physiologically meaningful patterns from ECG. This provides a promising foundation for enhancing clinical interpretability and supporting diagnostic decision-making, while also highlighting opportunities for further refinement of generative modelling approaches. Moreover, beyond direct clinical interpretation, these synthetic CMRs may serve as a feasible resource to enhance self-supervised pretraining or contrastive learning, thereby potentially extending the methodological utility of CardioNets. Additional generation results are available in the Supplementary Files.

**Reader Study**

To assess the performance of our AI model in a real-world clinical context, we conducted a reader study using an independent testing set of 111 subjects (77 cardiomyopathy patients and 34 controls, with paired ECG–CMR data) from The First Affiliated Hospital of Zhejiang University School of Medicine in 2024. The set was deliberately unselected, reflecting true clinical prevalence. The study consisted of two stages. In the first stage, six physicians (two resident physicians, two attending physicians, and two associate chief physicians) assessed each patient for cardiomyopathy using only ECG signals. Following a one-month washout period, the second stage involved re-evaluation with both real and AI-generated CMR images.

As shown in Fig. 6a, most physicians demonstrated limited screening performance when using ECG alone. The addition of AI-generated CMR notably improved diagnostic performance for resident and attending physicians. Across all participants, the combination of ECG and real CMR yielded the highest screening performance among human readers. However, it still fell short of CardioNets, which achieved the best performance with an AUC of 0.888. In FAHZU Dataset 3, we also compared CardioNets against a CMR reference model (CMR Ref) pretrained on UKB CMR data. CardioNets achieved a higher AUC (0.888) than the CMR Ref (0.840). We hypothesize that this difference arises because the CMR Ref relied on a CMR encoder pretrained on UKB, and heterogeneity in acquisition protocols and scanner settings between UKB and FAHZU likely limited its transferability, especially given the relatively small sample size. In contrast, ECGs are highly standardized across centers, with relatively uniform acquisition methods and vendor equipment, thereby minimizing confounding factors and making CardioNets more robust for cross-center opportunistic screening. Fig. 6b presents the average confusion matrices illustrating the ability of resident and attending physicians versus associate chief physicians to distinguish real from generated CMR images. The associate chief physicians more frequently misclassified real CMR images as AI-generated ones, likely due to their greater reliance on empirical experience and subjective interpretation - which, to some extent, demonstrates that the generated CMR images achieved such realistic quality that they could effectively confuse even expert judgment. Detailed confusion matrices for each individual physician are provided in Supplementary Fig. 9. In Fig. 6c, we compared the accuracy, sensitivity, specificity, and F1-score between all human physicians using ECG plus real CMR and CardioNets using ECG only. CardioNets demonstrated substantial improvements, achieving an accuracy 15.2% higher than the human average. Additionlly, the model inference is faster and more cost-effective than physician assessments, providing a scalable and economical solution for large-scale population screening.

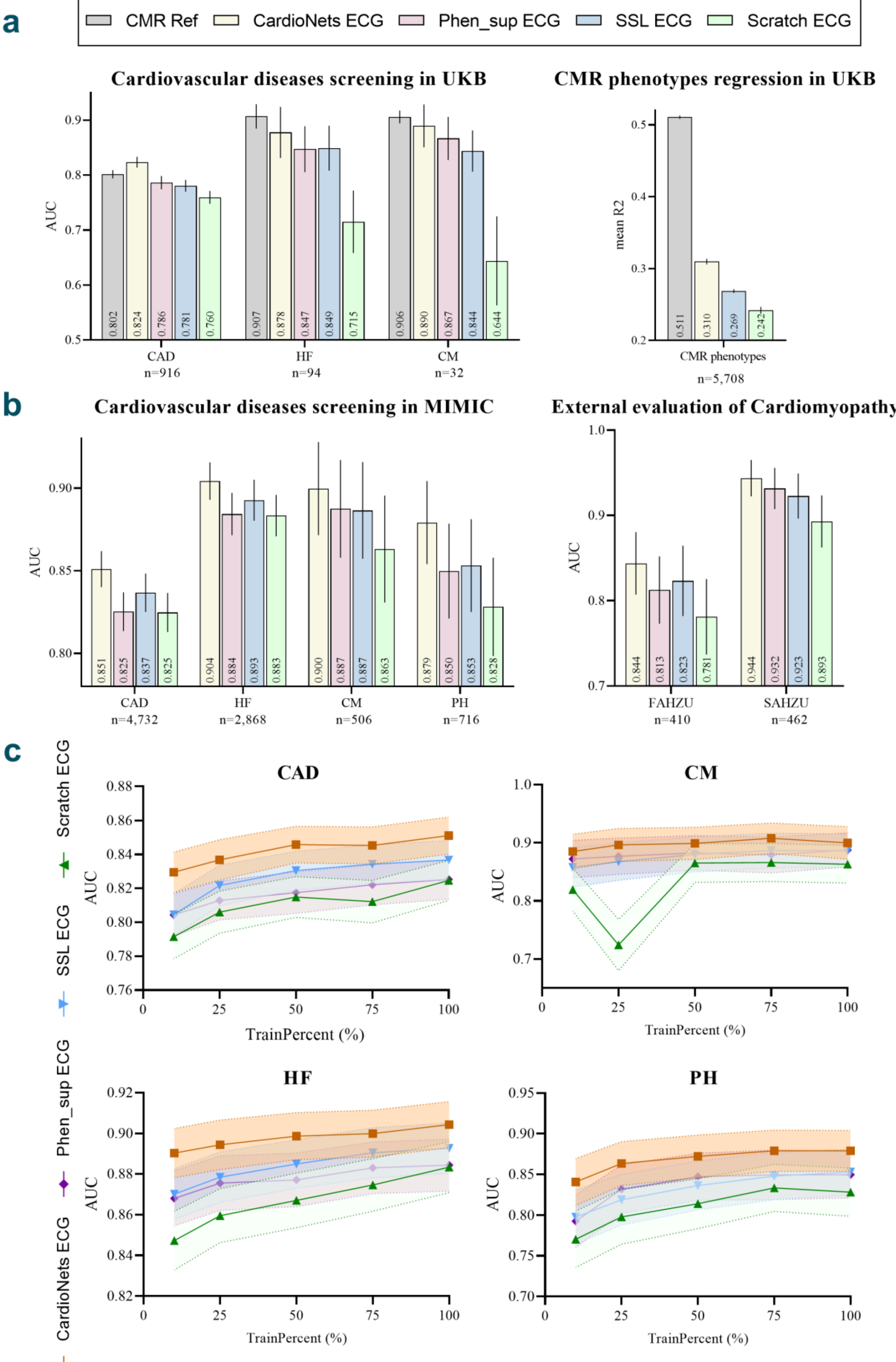

**Fig. 3: Overall assessment of cardiovascular status and label utilization efficiency. a.** The classification AUC metrics for coronary artery disease (CAD), cardiomyopathy (CM) and heart failure (HF), along with the average R2 values for cardiac structural phenotypes, were assessed in the UKB dataset. CardioNets was evaluated against four types of models: ECG models trained from scratch, ECG models fine-tuned from self-supervised (SSL) encoders, phenotype-supervised ECG models (Phen_sup ECG) pretrained on cardiac phenotype regression, and CMR-based models trained from SSL-pretrained encoders. Error bars represent the standard deviation across five-fold cross-validation. **b.** Screening of CVDs, including CAD, CM, HF, and pulmonary hypertension (PH), in the MIMIC dataset, and external validation for cardiomyopathy screening in FAHZU and SAHZU datasets. Due to the absence of corresponding CMR images in the MIMIC and private datasets, we compared CardioNets with ECG-based models. Error bars were computed using the bootstrap method with 1,000 resamples to estimate the 95% confidence interval (CI). Both panel a and panel b report n as the number of samples in the test set. **c.** Label efficiency was analysed by measuring the model's performance with varying fractions of training data, providing insight into the amount of data required to achieve target performance levels. This experiment was conducted on the MIMIC dataset using 10%, 25%, 50%, 75%, and 100% of the available training data. Error bars were derived using the same bootstrap procedure as in panel b (1,000 resamples, 95% CI), with shaded regions around the curves indicating the confidence intervals.

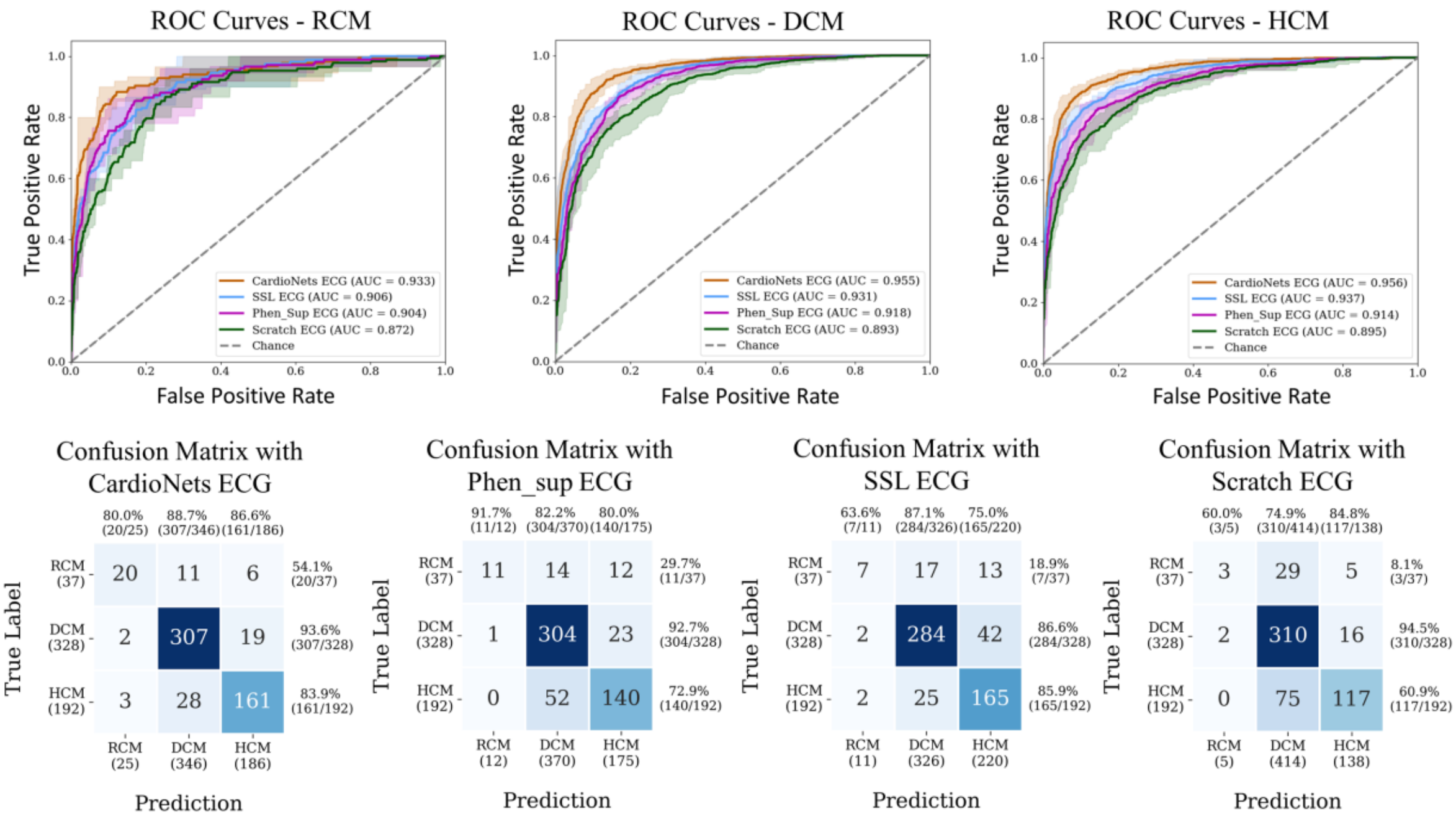

**Fig. 4: Classification results of the three subtypes of cardiomyopathy in FAHZU.** We compared CardioNets with three ECG-based baselines—SSL ECG, Phen_sup ECG, and Scratch ECG—in classifying restrictive cardiomyopathy (RCM), dilated cardiomyopathy (DCM), and hypertrophic cardiomyopathy (HCM). Results are presented as AUC curves using the one-vs-rest (OvR) approach for each subtype (up), along with confusion matrices for the three-class classification task across all methods (down).

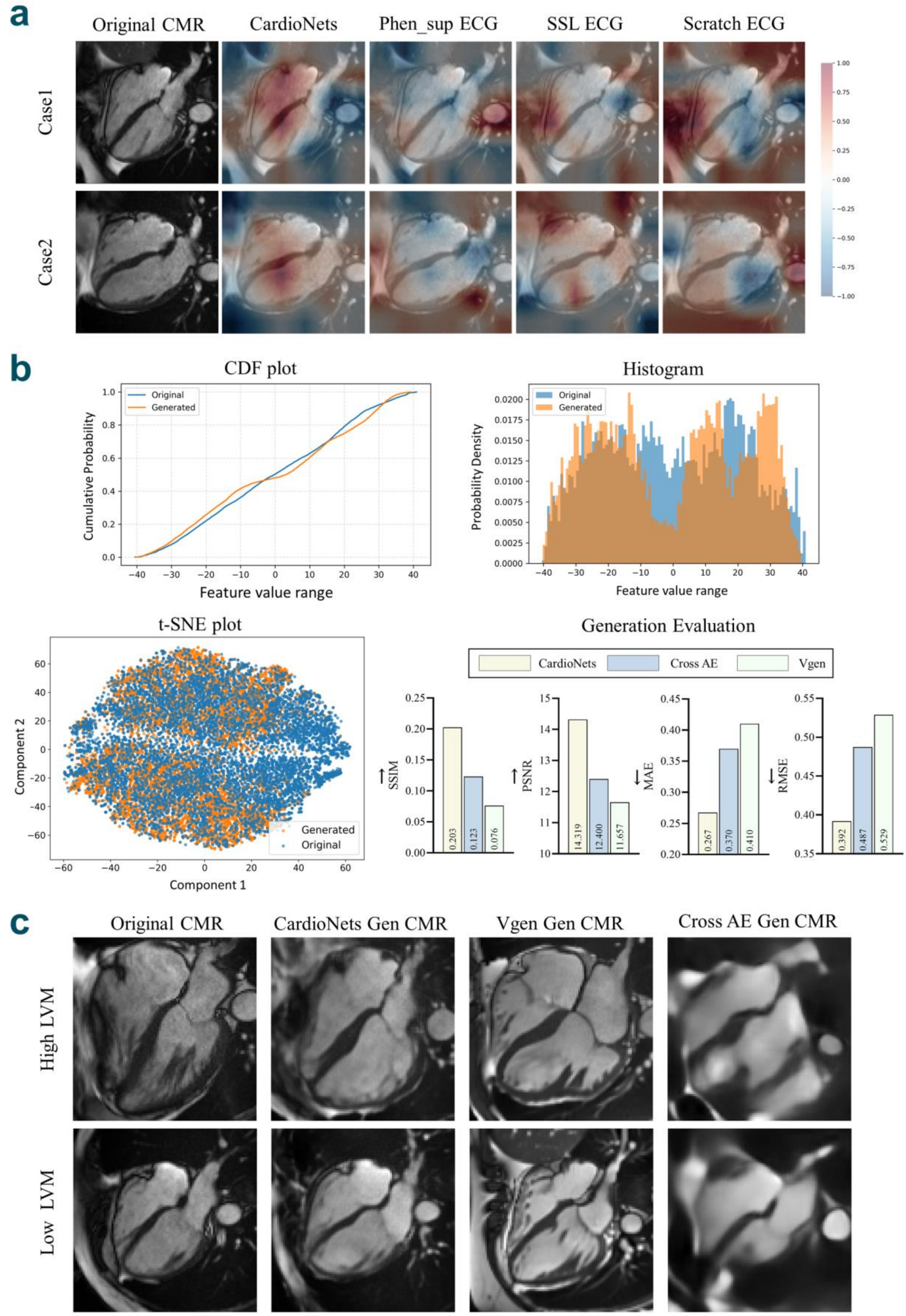

**Fig. 5: Interpretability heatmap of CMR and quality assessment of generated CMR images. a.** Cosine similarity heatmaps of CardioNets, SSL ECG, Phen_sup ECG, and Scratch ECG, CardioNets shows a greater interest in CMR cardiac regions. **b.** Comparison of feature consistency and paired quantitative metrics (SSIM,

PSNR, MAE, RMSE) between CardioNets-generated and real CMR images. **c.** Comparison of CMR images generated by CardioNets, Vgen[28], and Cross-AE[24] for cases with high and low left ventricular myocardial mass.

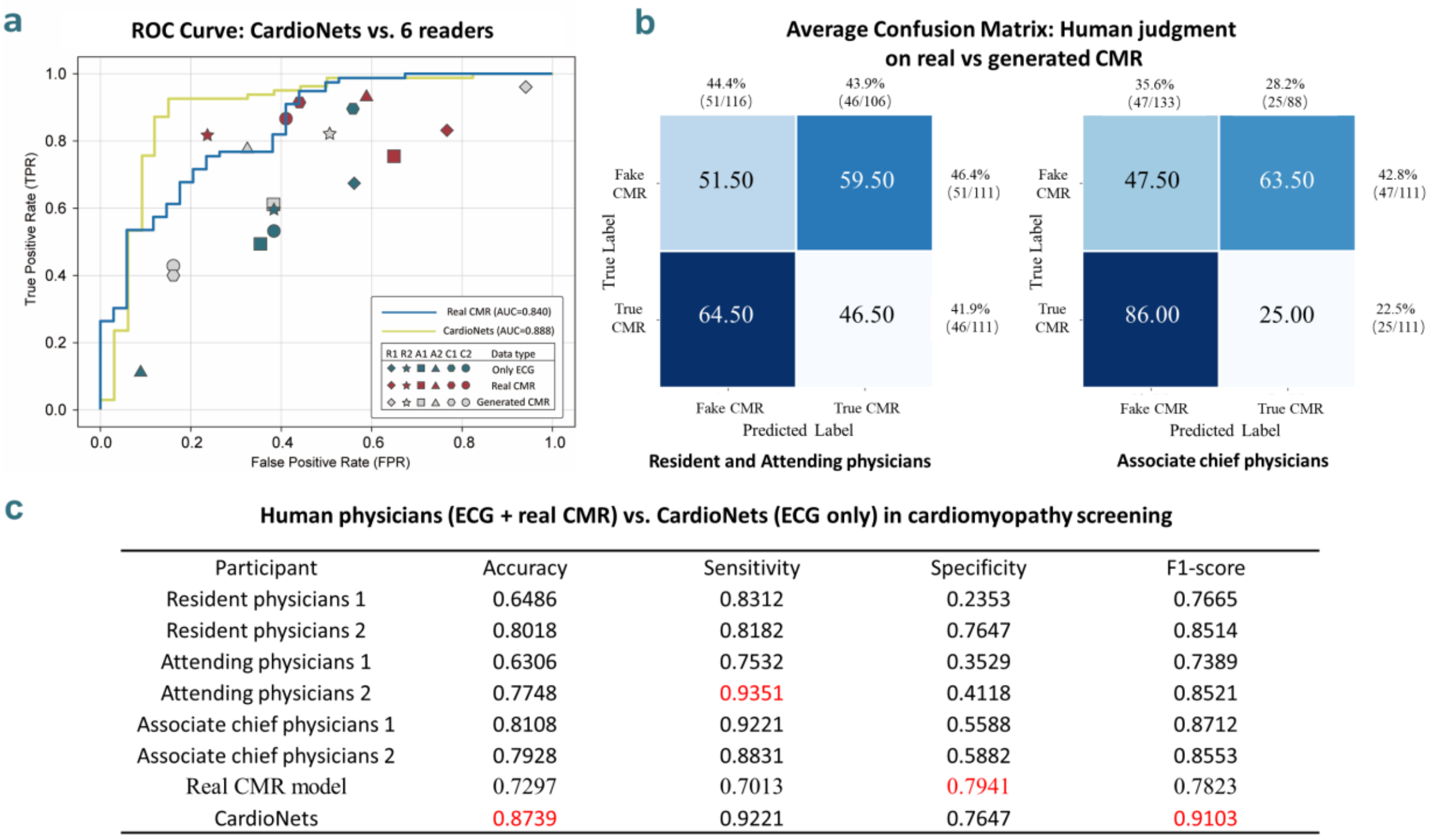

| Participant | Accuracy | Sensitivity | Specificity | F1-score |
| --- | --- | --- | --- | --- |
| Resident physicians 1 | 0.6486 | 0.8312 | 0.2353 | 0.7665 |
| Resident physicians 2 | 0.8018 | 0.8182 | 0.7647 | 0.8514 |
| Attending physicians 1 | 0.6306 | 0.7532 | 0.3529 | 0.7389 |
| Attending physicians 2 | 0.7748 | 0.9351 | 0.4118 | 0.8521 |
| Associate chief physicians 1 | 0.8108 | 0.9221 | 0.5588 | 0.8712 |
| Associate chief physicians 2 | 0.7928 | 0.8831 | 0.5882 | 0.8553 |
| Real CMR model | 0.7297 | 0.7013 | 0.7941 | 0.7823 |
| CardioNets | 0.8739 | 0.9221 | 0.7647 | 0.9103 |

**Fig. 6: Reader study. a.** Performance of AI models (CardioNets and a CMR-based reference model) and physicians in detecting cardiomyopathy across two stages: the first stage involved screening based solely on ECG interpretation, while the second stage provided physicians with either AI-generated or real CMR images. The reader study included two resident physicians (R1–R2), two attending physicians (A1–A2), and two associate chief physicians (C1–C2). **b.** Confusion matrix of physicians distinguishing generated from real CMR images. **c.** Comparison between human physicians using ECG and true CMR for cardiomyopathy screening, a CMR-based reference model (CMR Ref), and CardioNets using ECG alone.

## Discussion

In this study, we introduce CardioNets, a cross-modal pretraining model that leverages a pretraining-finetuning paradigm to efficiently adapt to the screening of a broad spectrum of CVDs. The visualizations generated by the model further enhance the interpretability of its results. CardioNets demonstrates substantial potential for utilizing ECG in CVD screening, particularly in economically underdeveloped regions where access to advanced diagnostic tools may be limited.

Building on our previous work[25], we recognized that prior strategies relying on supervised CMR models for task-specific alignment faced challenges in generalizability, as they required retraining both the CMR and ECG alignment models for each new task. To address this limitation, we developed CardioNets, which builds on self-supervised pretraining of a CMR encoder, freezing its parameters and aligning ECG representations through contrastive learning. As a result, the aligned ECG embeddings effectively capture holistic cardiac structural and functional information.

Although the UK Biobank (UKB) cohort offers a large number of paired ECG–CMR samples (42,483 in our study), the majority of participants are healthy individuals. Specifically, there are 2,289 cases of CAD (5.4%), 234 cases of HF (0.6%), and 81 cases of CM (0.2%). This imbalance raises concerns about the adequacy of disease representation. However, modern foundation models have demonstrated that large-scale pretraining[46-48], even on predominantly healthy or imbalanced data, can still yield robust downstream representations through the pretraining–finetuning paradigm. This approach has been demonstrated to mitigate long-tail issues and enable zero-shot learning across modalities and diseases. Similarly, despite the limited proportion of disease cases in UKB, large-scale pretraining within our framework remains beneficial for downstream cardiovascular tasks.

To further mitigate the imbalance and ensure disease-related features are adequately represented, we designed a dedicated strategy for CMR representation learning. First, we employed our phenotype-guided generative model (CPGG)[30] to augment the dataset by synthesizing five-fold additional CMRs, which, together with real CMRs, were used for self-supervised pretraining— a strategy also validated in other medical imaging studies[49]. This augmentation enriched the diversity of pathological phenotypes in the training set. In addition, during ECG–CMR contrastive learning, we implemented a weighted sampling strategy to minimize the impact of hard negatives and improve feature alignment. Together, these design choices enabled CardioNets to extract richer ECG embeddings that accurately capture structural cardiac information. The resulting embedding space encodes both healthy and diseased phenotypes, providing a robust and discriminative foundation for downstream analyses.

Substantial evidence shows that routine tests such as ECG or echocardiography offer limited diagnostic value for CVD detection[50,51], whereas CMR is considered the reference standard[52] but remains impractical for large-scale screening due to its complexity and cost. This raises the question of whether ECG, enriched with CMR-derived information, can serve as a more effective and accessible tool for CVD screening. The introduction of CardioNets demonstrates high performance across diverse downstream tasks, showing that the model can extract and represent comprehensive cardiac information—including structural features—from CMR using only ECG data. Our results in the UKB and MIMIC validate the potential of CardioNets for large-scale opportunistic screening. In addition, although the private datasets are modest in size, they were used solely for external validation under distribution shift rather than for training or hyperparameter selection. Despite their modest scale, these independent multi-center cohorts provide meaningful evidence of clinical generalizability. In the future, CardioNets may allow individuals to assess cardiovascular health and disease risk through routine examinations in simplified community hospital settings, thus reducing the risks associated with delayed diagnosis and treatment. This study builds on prior work[24] that mainly focused on representation learning of cardiac traits or global states, expanding the potential of ECG, when enriched with CMR-derived information, as a practical tool for population-level screening.

Our study demonstrates that ECG can not only screen for cardiomyopathy but also effectively identify its three subtypes. Cardiomyopathy is typically regarded as an irreversible heart disease[53], and distinct subtypes exhibit significant differences in cardiac structure and morphology[53]. These differences necessitate tailored treatment and prognostic approaches for each subtype[54]. Therefore, early screening of cardiomyopathy subtypes is crucial for supporting precision medicine and slowing disease progression. In clinical practice, precise identification of cardiomyopathy subtypes typically requires a comprehensive diagnostic workflow. The process usually begins with CMR screening to categorize subtypes[55], followed by a second step involving myocardial biopsy or genetic test[56] for biomarker extraction to confirm the diagnosis. Given the complexity of cardiomyopathy, the diagnostic process for its subtypes resembles a branching structure rather than a linear pathway[57], making CMR-based screening indispensable as a first step. Our proposed solution offers early probabilistic classification of cardiomyopathy subtypes using ECG alone. This finding highlights the potential for our model to replace CMR in subtype diagnosis, thereby streamlining the diagnostic process and conserving medical resources.

Interpretability is essential in clinical settings[58], Clinicians often emphasize the importance of visual evidence to support diagnostic reasoning, preferring interpretable structural information over abstract latent features. To address this need, CardioNets integrates a generative module that uses a masked autoregressive framework conditioned on ECG to produce temporal CMR images. These generated images provide physicians with intuitive visualizations that enhance both explainability and clinical confidence, even when actual CMR data are unavailable. In comparison to existing video diffusion models, such as Sora[59], which excel in creative domains, CardioNets is tailored specifically for medical applications. While creative generation focuses on artistic outputs, medical applications demand evidence-based results[60] and often work with relatively small datasets[61], making the direct application of general-purpose diffusion models impractical. By leveraging the aligned ECG encoder as a precise guide and training on a large-scale cross-modal dataset of ~40,000 ECG–CMR pairs, CardioNets generates high-resolution, high-fidelity CMR images that capture clinically meaningful structural phenotypes, outperforming previous methods. This dual capability—both predictive and visual—distinguishes CardioNets as not just a diagnostic aid, but also a potentially interpretable tool that can enhance clinical decision-making.

In the reader study, CardioNets, utilizing only ECG input, performed better to associate chief physicians evaluating both ECG and CMR, while significantly outperforming those who relied solely on ECG data. While our model achieves a favourable balance between specificity and sensitivity, as indicated by a high AUC, it is necessary to integrate the diagnostic logic of physicians into the model learning for enhancing human-machine interaction and fusion, potentially boosting the model's interpretability and clinical utility.

Several limitations must be considered. Firstly, while we validated the effectiveness of our solution using datasets from the UK, USA, and China, our pretrained model is based on approximately 40,000 ECG-CMR paired samples from the UK Biobank, which contains potential racial and socioeconomic biases[62]. Therefore, it is essential to retrain or update our model using a more diverse population and to conduct a thorough analysis of how well the model generalizes to underrepresented cohorts before applying it in clinical settings. Additionally, while the current CardioNets demonstrates capability in the coarse classification of the three principal cardiomyopathy categories, it exhibits limited discriminative power for specific subtypes, including but not limited to cardiac amyloidosis, Fabry disease, and glycogen storage disease. This highlights the need for further model refinement to enhance its performance across a broader range of cardiomyopathy subtypes. Secondly, while our model generates high-fidelity CMR from ECG, further research is needed to understand how ECG influences the generative model to generate accurate CMR images. Finally, prospective validation in diverse cohorts is needed to assess generalizability, and integrating clinician diagnostic logic into model training could enhance human-machine collaboration. Future work will explore real-world deployment in community hospitals to evaluate scalability and equity gains.

In conclusion, CardioNets demonstrates the potential of cross-modal learning for cardiac assessment using only ECG, providing a framework that may inform future strategies for disease screening at the population level. The model's successful application in the ECG-CMR domain also highlights its potential for analysing other paired datasets with strong-weak modality relationships, opening new avenues for multimodal research across diverse medical contexts.

## Competing Interests

The authors declare no competing interests.

## Acknowledgements

The authors thank all study participants and staff for contributing to the UK Biobank Cohort under application number 89757.

## Data availability

The data used in the study that supports cross-modal pre-training were obtained from the UK Biobank under application number 89757. Access to the UK Biobank data is available to all researchers with approval (https://www.ukbiobank.ac.uk/enable-your-research/register). The data used in the study that supports downstream tasks were obtained from the MIMIC-IV-ECG database, accessible via PhysioNet under credentialed access (https://physionet.org/content/mimic-iv-ecg/1.0/).

For the data from two private hospitals, the requirement for informed consent was waived by the respective ethics committees and institutions. The de-identified data can be shared only for non-commercial academic purposes and will require a formal material transfer agreement and a data use agreement. Requests should be submitted by emailing the corresponding authors (ZH, FW, HZ, or TC) at zhengxinghuang@zju.edu.cn, wufei@zju.edu.cn, doczhk@163.com, or ct010151452@zju.edu.cn. All requests will be evaluated based on institutional policies to determine whether the requested data are subject to intellectual property or patient privacy obligations. Generally, all such requests for access to data will be responded to within 1 month.

## Code availability

The source codes about this study and data analysis in this manuscript are provided at https://github.com/Yukui-1999/ECG-CMR.

## Funding

This work was partially supported by the National Key Research and Development Program of China (Grant No. 2022YFF1202400), and was further supported by the Technical Innovation Key Project of Zhejiang Province (2024C03023) and the National Natural Science Foundation of China (Grant Nos. 82272129, 82470428).

## Author contributions

ZH, HZ and FW jointly supervised research. ZH designed this study. ZD, ZY Li, and YH developed a deep learning model and performed the model interpretation. ZD, ZY Li, YM and HL performed data analysis. ZD, YH, ZK Li, WP, ZY Liu, XC, FW, TC, HZ and ZH interpreted the results. TC, QL, JW,

YC, MC, LW, YX and CZ participated in the reader study. ZD, ZY Li, YH, TC and ZH prepared the first draft of the manuscript. All authors contributed and approved the final draft.

## Extended Fig

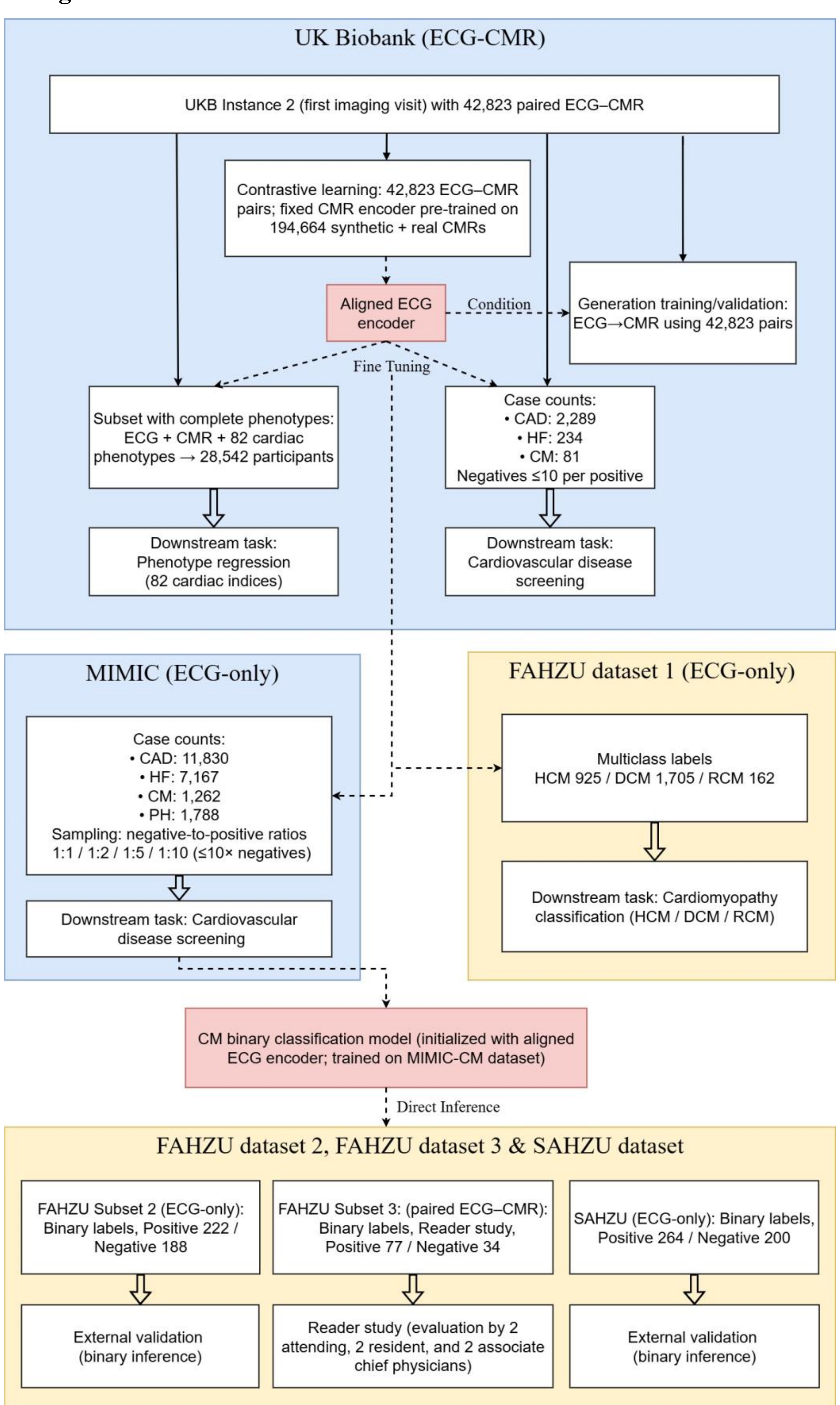

**Extended Fig. 1: Overall study workflow across datasets.** The workflow illustrates model pretraining on UKB ECG–CMR pairs, phenotype regression, cardiovascular disease screening in UKB and MIMIC, external validation in FAHZU and SAHZU, and a reader study. Detailed datasets characteristics, sample sizes, and exclusion criteria are provided in the Methods. Solid arrows indicate data selection and construction, dashed arrows denote model training and usage, and hollow arrows specify downstream tasks. Blue boxes represent public datasets, and yellow boxes represent private datasets.

**Fig-1: Overview of the study design. Stage I:** Development of CardioNets. Using the UKB dataset, we trained two core components of CardioNets: (1) an ECG model trained with ECG-CMR contrastive learning to capture relevant information from CMR, and (2) a CMR generation model trained with a masked autoregressive model conditioned on ECG. **Stage II:** Downstream tasks. The datasets used are from UKB, MIMIC-IV-ECG, The First Affiliated Hospital of Zhejiang University School of Medicine (FAHZU), The Second Affiliated Hospital of Zhejiang University School of Medicine (SAHZU). All numbers annotated in the figure are represent the total sample sizes of the datasets. All tasks used only ECG data to validate the model's performance, including: (A) Prediction of cardiac phenotypes, (B) Screening for cardiovascular diseases, particularly cardiomyopathy, (C) Interpretability study of ECG-CMR align and evaluation of CMR generation, and (D) Reader study for cardiomyopathy screening.

**Fig-1** Source line: Introduction, paragraph 3, line 2.

**Fig. 2: Model diagram. a.** CMR self-supervised model, leveraging masked self-supervised learning and optimized to minimize reconstruction loss. To enrich pretraining, we incorporated a phenotype-guided CMR generation approach (from our previous work[30]), which expanded the dataset with mixed real and generated CMRs, thereby enabling the model to learn more robust CMR representations. **b.** ECG-CMR contrastive learning, with the parameters of the CMR self-supervised model in subgraph a frozen, optimized to minimize the weighted contrastive loss, where $w_{i,j}$ denotes the phenotype-based weight and $s_{i,j}$ denotes the cosine similarity between embeddings. **c.** ECG-guided CMR image generation, using the pre-trained ECG encoder in subgraph b as the feature extractor. **d.** Fine-tuning the aligned ECG encoder to adapt to various downstream tasks and use pretrained generative model in subgraph c generate the corresponding CMR images.

**Fig-2a** Source line: Method—Self-supervised method for CMR, paragraph 1, line 1.

**Fig-2b** Source line: Method—Contrastive ECG-CMR Pre-training, paragraph 1, line 3.

**Fig-2c** Source line: Method—ECG2CMR generative model, paragraph 3, line 2.

**Fig-2d** Source line: Method—Downstream tasks, paragraph 1, line 2.

**Fig. 3: Overall assessment of cardiovascular status and label utilization efficiency**. **a**. The classification AUC metrics for coronary artery disease (CAD), cardiomyopathy (CM) and heart failure (HF), along with the average R2 values for cardiac structural phenotypes, were assessed in the UKB dataset. CardioNets was evaluated against four types of models: ECG models trained from scratch, ECG models fine-tuned from self-supervised (SSL) encoders, phenotype-supervised ECG models (Phen_sup ECG) pretrained on cardiac phenotype regression, and CMR-based models trained from SSL-pretrained encoders. Error bars represent the standard deviation across five-fold cross-validation. **b**. Screening of CVDs, including CAD, CM, HF, and pulmonary hypertension (PH), in the MIMIC dataset, and external validation for cardiomyopathy screening in FAHZU and SAHZU datasets. Due to the absence of corresponding CMR images in the MIMIC and private datasets, we compared CardioNets with ECG-based models. Error bars were computed using the bootstrap method with 1,000 resamples to estimate the 95% confidence interval (CI). Both panel a and panel b report n as the number of samples in the test set. **c**. Label efficiency was analysed by measuring the model's performance with varying fractions of training data, providing insight into the amount of data required to achieve target performance levels. This experiment was conducted on the MIMIC dataset using 10%, 25%, 50%, 75%, and 100% of the available training data. Error bars were derived using the same bootstrap procedure as in panel b (1,000 resamples, 95% CI), with shaded regions around the curves indicating the confidence intervals.

**Fig-3a-b** Source line: Results—Cardiac assessments, paragraph 1, line 3.

**Fig-3c** Source line: Results—Cardiac assessments, paragraph 5, line 4.

**Fig-4: Classification results of the three subtypes of cardiomyopathy in FAHZU.** We compared CardioNets with three ECG-based baselines—SSL ECG, Phen_sup ECG, and Scratch ECG—in classifying restrictive cardiomyopathy (RCM), dilated cardiomyopathy (DCM), and hypertrophic cardiomyopathy (HCM). Results are presented as AUC curves using the one-vs-rest (OvR) approach for each subtype (up), along with confusion matrices for the three-class classification task across all methods (down).

**Fig-4** Source line: Results—Screening for subtypes of cardiomyopathy, paragraph 1, line 3.

**Fig-5: Interpretability heatmap of CMR and quality assessment of generated CMR images. a**. Cosine similarity heatmaps of CardioNets, SSL ECG, Phen_sup ECG, and Scratch ECG, CardioNets shows a greater interest in CMR cardiac regions. **b.** Comparison of feature consistency and paired quantitative metrics (SSIM, PSNR, MAE, RMSE) between CardioNets-generated and real CMR images. **c.** Comparison of CMR images generated by CardioNets, Vgen[28], and Cross-AE[24] for cases with high and low left ventricular myocardial mass.

**Fig-5a** Source line: Results—Interpretation study and evaluation of CMR generation, paragraph 1, line 2.

**Fig-5b-c** Source line: Results—Interpretation study and evaluation of CMR generation, paragraph 2, line 2.

**Fig-6: Reader study. a.** Performance of AI models (CardioNets and a CMR-based reference model) and physicians in detecting cardiomyopathy across two stages: the first stage involved screening based solely on ECG interpretation, while the second stage provided physicians with either AI-generated or real CMR images. The reader study included two resident physicians (R1–R2), two attending physicians (A1–A2), and two associate chief physicians (C1–C2). **b.** Confusion matrix of physicians distinguishing generated from real CMR images. **c.** Comparison between human physicians using ECG and true CMR for cardiomyopathy screening, a CMR-based reference model (CMR Ref), and CardioNets using ECG alone.

**Fig-6a** Source line: Results—Reader Study, paragraph 2, line 1.

**Fig-6b** Source line: Results—Reader Study, paragraph 2, line 12.

**Fig-6c** Source line: Results—Reader Study, paragraph 2, line 17.

**Extended-Fig-1: Overall study workflow across datasets.** The workflow illustrates model pretraining on UKB ECG–CMR pairs, phenotype regression, cardiovascular disease screening in UKB and MIMIC, external validation in FAHZU and SAHZU, and a reader study. Detailed datasets characteristics, sample sizes, and exclusion criteria are provided in the Methods. Solid arrows indicate data selection and construction, dashed arrows denote model training and usage, and hollow arrows specify downstream tasks. Blue boxes represent public datasets, and yellow boxes represent private datasets.

**Extended-Fig-1** Source line: Method—Dataset for Model Development and Evaluation, paragraph 1, line 2.